%% file: apstemplate.tex
\documentclass[aps,prd,reprint,preprintnumbers,superscriptaddress,showpacs,floatfix,nofootinbib]{revtex4-2}
\usepackage[utf8]{inputenc}
\usepackage{parskip}
\usepackage{amssymb}
\usepackage{hhline}
\usepackage{amsmath}
\usepackage{bm}
\usepackage[dvipsnames]{xcolor}
\usepackage{xspace}
\usepackage{multirow,tabularx}
\usepackage{siunitx}
\usepackage{textgreek}
\usepackage{upgreek}
\usepackage{multirow}
\usepackage{graphicx}
\usepackage{pifont}
\usepackage{xstring}
\usepackage{etoolbox}
\usepackage{stmaryrd}
\usepackage{natbib}

\DeclareSIUnit\parsec{pc} 
\DeclareSIUnit\littleh{\mathsf{h}} 
\DeclareSIUnit\ccs{{m_{\text{p}}}^2} 
\DeclareSIUnit\nothing{\relax} 

\def\araa{\rm{ARA\&A}}
\def\apj{\rm{ApJ}}

\def\mnras{\rm{MNRAS}}

\def\nar{\rm{New A Rev.}}

\def\prd{\rm{Phys.~Rev.~D}}

\newcommand{\aneq}{\mathrel{\smash[t]{\stackrel{{\text{an}}}{=}}}} 

\input{cosmic_class_names.tex} 

\usepackage{hyperref}
\hypersetup{
     colorlinks = true,
     linkcolor = Blue,
     citecolor = Blue,
     filecolor = Blue,
     urlcolor = Blue 
     }
\usepackage[capitalize]{cleveref} 

\begin{document}

\title{Mapping Poincar\'e gauge cosmology to Horndeski theory for emergent dark energy} 

\author{W.E.V. Barker}
\email{wb263@cam.ac.uk}
\affiliation{Astrophysics Group, Cavendish Laboratory, JJ Thomson Avenue, Cambridge CB3 0HE, UK}
\affiliation{Kavli Institute for Cosmology, Madingley Road, Cambridge CB3 0HA, UK}
\author{A.N. Lasenby}
\email{a.n.lasenby@mrao.cam.ac.uk}
\affiliation{Astrophysics Group, Cavendish Laboratory, JJ Thomson Avenue, Cambridge CB3 0HE, UK}
\affiliation{Kavli Institute for Cosmology, Madingley Road, Cambridge CB3 0HA, UK}
\author{M.P. Hobson}
\email{mph@mrao.cam.ac.uk}
\affiliation{Astrophysics Group, Cavendish Laboratory, JJ Thomson Avenue, Cambridge CB3 0HE, UK}
\author{W.J. Handley}
\email{wh260@cam.ac.uk}
\affiliation{Astrophysics Group, Cavendish Laboratory, JJ Thomson Avenue, Cambridge CB3 0HE, UK}
\affiliation{Kavli Institute for Cosmology, Madingley Road, Cambridge CB3 0HA, UK}


\begin{abstract}
  The ten-parameter, quadratic Poincar\'e gauge theory of gravity is a plausible alternative to general relativity. We show that the rich background cosmology of the gauge theory is described by a non-canonical bi-scalar-tensor theory in the Jordan frame: the \emph{metrical analogue}. This provides a unified framework for future investigation by the broader community. For many parameter choices, the non-canonical term reduces to a \emph{Cuscuton} field of the form $\smash{\sqrt{|X^{\phi\phi}|}}$. The Einstein--Cartan--Kibble--Sciama theory maps to a pure quadratic cuscuton, whereas the teleparallel theory maps to the Einstein--Hilbert Lagrangian. We apply the metrical analogue to novel unitary and power-counting-renormalisable cases of Poincar\'e gauge theory. These theories support the concordance \textLambda CDM background cosmology up to an optional, effective dark radiation component, we explain this behaviour in terms of a stalled cuscuton. We also obtain two dark energy solutions from one of these cases: accelerated expansion from a \emph{negative} bare cosmological constant whose magnitude is screened, and emergent dark energy to replace \emph{vanishing} bare cosmological constant in \textLambda CDM.
\end{abstract}


\maketitle

\section{Introduction}\label{introduction}
Candidate discrepancies between the cosmic concordance model (\textLambda CDM)~\cite{2016PDU....12...56B,2018arXiv180401318S} and observation~\cite{2019NatAs...3..891V,2019arXiv190809139H,2019arXiv191102087D,2017ARA&A..55..343B,2019NatRP...2...10R,2019ApJ...876...85R,2018arXiv180706209P} have fuelled interest in modifications to general relativity (GR). In order to bypass Lovelock's theorem, scalar-tensor theories couple various scalar fields $\phi$ to the metric $g_{\mu\nu}$ on a curved spacetime $\mathcal{  M}$~\cite{2019IJMPD..2830012Q}. This approach is prevalent in effective field theory (EFT) extensions to GR, and even used to model inflation within \textLambda CDM~\cite{2008JHEP...03..014C}. Scalar-tensor theories are tractable and very widely studied, and in this sense they are \emph{self-motivating}. 

The EFT approach to gravity is motivated in part by the perturbative non-renormalisability of GR, in which the superficial divergence of a diagram scales with the number of loops~\cite{Buoninfante:2016iuf}. Even at one loop, the inclusion of matter propagators spoils the renormalisability of pure GR by invoking quadratic curvature counter-terms: such terms cannot be absorbed into the linear curvature invariant by rescaling \cite{1974AIHPA..20...69T,1974PhRvD..10..401D,1974PhRvD..10.3337D}. A possible solution is to add such terms to the Einstein--Hilbert Lagrangian a priori. This approach culminates in the renormalisable theory of Stelle \cite{1977PhRvD..16..953S}. The addition of quadratic curvature invariants motivated in the ultraviolet (UV) should not interfere with the usual tests passed by GR in the infrared (IR). However, they necessarily result in higher derivative theories whose unitarity may be questionable under standard quantisation schemes. For example, Stelle's theory contains a ghost in its tree-level graviton propagator, although recently this has been argued not to prevent unitary at the QFT level~\cite{Larin:2019zhb}. 

The problematic link between quadratic curvature additions and higher derivatives may be broken by reconsidering the fundamental dynamical variables of gravity. This route was suggested already by gravitational coupling to spinors, which requires $g_{\mu\nu}$ to be split into tetrad (or vierbein) fields. Higher derivatives still persist if the (spin) connection is expressed in terms of tetrad derivatives. The required leap is then to treat the spin connection as a separate dynamical field \emph{at all times}, arriving at the Poincar\'e gauge theory (PGT) of Kibble~\cite{1961JMP.....2..212K}, Utiyama~\cite{PhysRev.101.1597}, Sciama~\cite{RevModPhys.36.463} and others. 
The particular interpretation of PGT employed in this paper is set in a flat spacetime $\check{\mathcal{  M}}$, in which the tetrad and spin connection are cast as translational and rotational gauge fields $h_a^{\ \mu}$ and $\smash{A^{ab}_{\ \ \mu}\equiv A^{\smash{[}ab\smash{]}}_{\ \ \mu}}$.

The minimal PGT extension to GR is usually taken to be Einstein--Cartan--Kibble--Sciama (ECKS) theory~\cite{2006gr.qc.....6062T}, which has an Einstein--Hilbert Lagrangian. ECKS theory is dynamically equivalent to GR in the absence of fermionic matter, but otherwise admits \emph{torsion} in contact with matter spin sources. PGT enjoys a measure of `naturalness' because torsion is inherently allowed, though it may vanish dynamically as in ECKS theory. In GR, torsion is artificically suppressed at the level of the covariant formulation. While torsion has not been directly observed, we note the relative paucity of theories in which such an observation is expected or even possible~\cite{2010RPPh...73e6901N,1997PhLA..228..223L,2019arXiv190304712B,2020EPJC...80..559C}, among them theories with highly non-minimal matter couplings~\cite{2014IJMPD..2342004P}. Of greater concern is the \emph{algebraic} nature of the spin-torsion interaction particular to ECKS theory. For more natural \emph{dynamical} torsion, one again requires quadratic curvature additions to the Einstein--Hilbert Lagrangian. However, it is precisely the addition of quadratic invariants which further challenges the position of ECKS theory as the minimal PGT extension to GR. The \emph{teleparallel} Lagrangian is a specific linear combination of the three quadratic torsion invariants~\cite{2018arXiv180106929G}. Under the assumption of vanishing curvature (which may be achieved via multiplier fields~\cite{blagojevic2002gravitation}), this theory is \emph{also} dynamically equivalent to GR. We note that in the more general metric-affine gauge theory (MAGT), a similar analogue exists for quadratic non-metricity invariants under the assumption of vanishing curvature and torsion: this theory is known as \emph{coincident GR} (CGR)~\cite{2017arXiv171003116B,2019Univ....5..173B}.

The present paper is restricted to PGT, in which, in addition to the Einstein--Hilbert term, there are six quadratic curvature invariants\footnote{Typically only five quadratic curvature invariants are considered to be independent due to the Gauss--Bonnet identity~\cite{Hayashi:1980bf}.}, and no reason to exclude the further three quadratic torsion invariants. Extra quadratic invariants may only be formed at the cost of parity-violating Lagrangia~\cite{2018PhRvD..98b4014B}. Consequently, many authors have considered the \emph{ten-parameter}, \emph{quadratic}, \emph{parity-preserving} PGT (PGT\textsuperscript{q,+}).
The inhomogeneous transformation of $A^{ab}_{\ \ \mu}$ under external Lorentz rotations, all set on a flat background, bears a strong resemblance to the Yang--Mills theories of internal isospin symmetries. In this sense, a Lagrangian quadratic in the field strength tensors comes somewhat more naturally to PGT than to its metrical counterpart. Moreover, the Yang--Mills analogy seems further to reinforce our initial guess that a perturbative approach to renormalisation \emph{could} be taken. Popular alternatives exist, and some of these have Yang--Mills counterparts, too. For example, the asymptotic safety of QCD \cite{2005RvMP...77..837G} has helped to drive the quest for fixed points in the renormalisation group flow~\cite{2018FrASS...5...47E}.

Building on early work by Neville~\cite{1978PhRvD..18.3535N,1980PhRvD..21..867N}, Sezgin and van Nieuwenhuizen identified $12$ cases of pure PGT\textsuperscript{q,+} which appear unitary at the level of their linearised Lagrangia~\cite{1980PhRvD..21.3269S,1981PhRvD..24.1677S}.
In the linearised picture, propagating $h_a^{\ \mu}$ and $A^{ab}_{\ \ \mu}$ modes are respectively termed \emph{gravitons} and \emph{rotons} (or \emph{tordions}). PGT\textsuperscript{q,+} admits a massless graviton of spin-parity $J^P=2^+$, and six rotons which may be $0^{\pm}$, $1^{\pm}$ or $2^{\pm}$, and which have (generally nonzero) mass parameters. While $h_a^{\ \mu}$ and $A^{ab}_{\ \ \mu}$ provide $16+24$ degrees of freedom (D.o.F), Poincar\'e gauge symmetry eliminates $2\times 10$ D.o.F, and the remaining $20$ D.o.F are accounted for by these modes.
Unitarity is achieved by eliminating propagator poles with negative residues and imaginary masses. 
Quite recently, $58$ additional cases were discovered by~\citet*{2019PhRvD..99f4001L,2020PhRvD.101f4038L}: these are not only unitary, but also power-counting renormalisable (PCR).
The `strong' PCR criteria stipulate that the graviton propagator should approach the UV as $p^{-4}$ (in common with the quadratic metrical theory), while the roton propagator should tend to $p^{-2}$ due to the extra momentum dependence of its vertices \cite{1978PhRvD..18.3535N,2016JMP....57i2505L}. In all but $4$ of the $58$ novel cases, modes exist which violate the strong criteria: these decouple in the UV and produce no divergent loops~\cite{2020PhRvD.101f4038L}.
Superficially therefore, PGT\textsuperscript{q,+} does appear to reward the expectation of simultaneous unitarity and renormalisability. However, there are several important caveats.

Principally, power counting is only a \emph{proxy} for perturbative renormalisability: neither the Ward--Takahashi identities nor the primitively divergent diagrams have been identified. Moreover, no allowance is made for matter loops, which are already known to hinder the perturbative approach to GR~\cite{1974AIHPA..20...69T,1974PhRvD..10..401D,1974PhRvD..10.3337D}. Most importantly, the $12$ original unitary cases and $58$ novel unitary and PCR cases were proposed by considerations in the \emph{weak field regime}~\cite{1980PThPh..64.1435H}. Yo and Nester applied the Dirac--Bergmann algorithm (schematically) to the initial $12$ cases: most of them fall apart in the strong regime, so that strictly ghostly sectors may become excited~\cite{2002IJMPD..11..747Y}. Their analysis also gave indications of tachyonic instability, albeit through less well established methods~\cite{1998AcPPB..29..961C}. This followed their earlier (complete) implementation of the algorithm for a few cases with only massive $0^+$ or $0^-$ rotons in addition to the massless $2^+$ graviton, with favourable results~\cite{1999IJMPD...8..459Y}. In fact, not only the initial $12$ cases but also the great preponderance of theories in the literature feature the massless $2^+$ graviton \emph{automatically} by including the Einstein--Hilbert term. This term is missing from all the $58$ novel cases: as with teleparallelism and CGR, these are \emph{exclusively quadratic theories}. 

It is not yet clear whether the loss of the linear curvature invariant is a `feature' or a `bug'. Of the $58$ novel cases, $19$ contain two massless D.o.F, of which only $17$ have a nonvanishing $2^+$ propagator. For these cases the massless roton may be tentatively identified with the unique `graviton' of Weinberg and Witten~\cite{Porrati:2012rd}, so that the theory may still be viable. In fact, all $58$ novel cases feature vanishing roton mass parameters, but the relevant $J^P$ sectors are usually non-propagating. This is another point of contrast with the literature: vanishing mass parameters are typically avoided due to their association with emergent gauge symmetries, which in turn drastically complicate the Hamiltonian constraint chain~\cite{1983PhRvD..28.2455B,1987PhRvD..35.3748B}. For this and other reasons, we note that the $58$ novel cases are (for the moment) insulated from the earlier Hamiltonian surveys in~\cite{1999IJMPD...8..459Y,2002IJMPD..11..747Y}: a dedicated analysis is in preparation and will be presented in a companion paper~\cite{barker4}. Without the linear curvature invariant, we also lose contact with the established IR limit of the theory. On the one hand, we might embrace this as an unusual opportunity to motivate the IR from the UV, since the quadratic invariants are no longer perturbative corrections to the theory. On the other, we may view this as a potentially fatal flaw which necessitates disparate IR investigations into nature's observed spacetimes. 

The present paper is concerned with the spatially flat Friedmann--Lema\^itre--Robertson--Walker (FLRW) spacetime, which is a central axiom of the cosmological constant + cold dark matter model~\cite{2016PDU....12...56B,2018arXiv180401318S}. 
Recently, we used the homogeneity and isotropy of the strong cosmological principle (SCP) to partition a select $33$ of the $58$ novel cases into phenomenological \emph{classes}~\cite{2020arXiv200302690B}. The \emph{\cosmicclass{null}} theory reproduces the \textLambda CDM background. Moreover, an early-time deviation from \textLambda CDM dilutes away as \emph{dark radiation}, qualitatively suited to ease the present tension~\cite{2019NatRP...2...10R,2020arXiv200306396Z} between CMB-inferred $\SI[separate-uncertainty=true,multi-part-units=repeat]{0.674(5)}{\nothing}$~\cite{2018arXiv180706209P} and locally-observed $\SI[separate-uncertainty=true,multi-part-units=repeat]{0.735(14)}{\nothing}$~\cite{2019ApJ...876...85R} determinations of the contemporary Hubble number ${\mathsf{h}=H_0/\SI{100}{\kilo\metre\per\second\per\mega\parsec}}$. The more general \emph{\cosmicclass{cnull}} has an additional massive $0^-$ D.o.F, but is \emph{hitherto unexplored}. Separately, we emphasise that the cases underlying these classes simultaneously contain two massless (possibly $2^+$) D.o.F, pass basic Solar System tests and support the usual gravitational wave polarisations~\cite{Las}. Notwithstanding our earlier analysis, the cosmological equations of PGT\textsuperscript{q,+} are quite cumbersome and opaque. This has led to fruitful, but often piecewise investigations for almost forty years (see e.g.~\cite{1984CQGra...1..651G,2008PhRvD..78b3522S,2013JCAP...03..040M,2019arXiv190403545Z} or reviews of the substantial literature~\cite{2005NewAR..49...59P} and in~\cite{2020arXiv200302690B}).

The first aim of this paper is to develop a simple bi-scalar-tensor theory -- the \emph{metrical analogue} (MA) -- which reproduces the spatially-flat background cosmology of PGT\textsuperscript{q,+}. The general MA will be given in \cref{final} and provides a unified framework for future IR investigation by the broader community. Since the MA is free of both torsion and quadratic curvature invariants, we find that it offers a refreshingly clear statement of the IR. Just as $h_a^{\ \mu}$ is in some sense the \emph{square root} of $g_{\mu\nu}$, the MA contains a non-canonical kinetic term of the form $\smash{\sqrt{|X^{\phi\phi}|}}$, where $\smash{X^{\phi\phi}\equiv\tfrac{1}{2}g^{\mu\nu}\partial_\mu\phi\partial_\nu\phi}$. Such fields are known in cosmology as \emph{Cuscutons}~\cite{2007PhRvD..75h3513A}: they provide a rich phenomenology~\cite{2007PhRvD..75l3509A}, but are naturally challenging to motivate (see e.g. EFT applications in Ho\v{r}ava--Lifshitz gravity~\cite{2009PhRvD..80h1502A}). We will show that teleparallelism has an Einstein--Hilbert MA, while the MA of ECKS theory is a pure \emph{Cuscuton}. 

The second aim of this paper is to use the MA to study the IR of certain novel cases, which were partly motivated in the UV. We will show that \cosmicclass{cnull} of PGT\textsuperscript{q,+} inherits the dark radiation of \cosmicclass{null}, while the $0^-$ mass generates \emph{dark energy}. The \emph{Cuscuton} tends to `stall' the cosmology in a state equivalent to \textLambda CDM. With relevance to the Hubble tension and cosmological constant problem~\cite{2012CRPhy..13..566M,2011arXiv1105.6296K}, our results build the case for further careful scrutiny of the underlying novel cases.

The remainder of this paper is set out as follows. In \cref{metric_theories,tetrad_theories,scale_invariance} we map PGT\textsuperscript{q,+} to the MA. In \cref{application_to_novel_theories} we provide a brief primer on the novel theories of interest, and their potential for renormalisability. In \cref{negative_screened_dark_energy} we show that \cosmicclass{cnull} can undergo accelerated expansion in the presence of a \emph{negative} bare cosmological constant. In \cref{generally_viable_dark_energy} we find an alternative solution in which \textLambda CDM is recovered, but the cosmological constant is provided entirely by the gravitational sector. Conclusions follow in \cref{discussion}.
We use natural units $c\equiv\hbar\equiv 1$, reduced Planck mass ${m_{\text{p}}}^2\equiv\kappa^{-1}$ and signature $(+,-,-,-)$.

\section{Metric theories}\label{metric_theories}
The generalised galileon, more commonly known as Horndeski theory~\cite{1974IJTP...10..363H}, is the most general $\phi$--$g_{\mu\nu}$ coupling with maximally second-order field equations. Avoidance of higher-order field equations is a simple (yet insufficient) precaution against ghosts given by Ostrogradsky's theorem. The generalised bi-galileon~\cite{2013JHEP...04..032P} introduces a second scalar $\psi$ and is known \emph{not} to be the most general second-order bi-scalar-tensor theory~\cite{2013PhRvD..88h3504K}, but follows a simple prescription and is also often called Horndeski theory. The generality of the bi-galileon is provided by six arbitrary $G$-functions. Of these, it suits our needs to discard $\smash{G_3^{\phi}}$, $\smash{G_3^{\psi}}$, $\smash{G_5^{\phi}}$ and $\smash{G_5^{\psi}}$ (adopting the usual notation~\cite{2019RPPh...82h6901K}) for a total Lagrangian
\begin{equation}
  L_{\text{T}}= G_2^{\phantom{\psi}}(\phi,\psi;X^{\phi\phi},X^{\psi\psi})+G_4^{\phantom{\psi}}(\phi,\psi)R+L_{\text{m}}(\Phi;g).
  \label{horndeski_full_lagrangian}
\end{equation}
Note that $G_2^{\phantom{\psi}}$ couples $\partial\phi$ and $\partial\psi$ to $g_{\mu\nu}$, and $G_4^{\phantom{\psi}}$ non-minimally couples $\phi$ and $\psi$ to $\partial g$ and $\partial^2g$ via the Ricci scalar $R\equiv R^{\mu\nu}_{\ \ \mu\nu}$, where the Riemann tensor is
\begin{equation}
  R_{\alpha\beta\mu}^{\ \ \ \ \nu}\equiv 2\big(\partial^{\phantom{\lambda}}_{[\beta}\Gamma^{\nu\phantom{\lambda}}_{\ \alpha]\mu}+\Gamma^\lambda_{\ [\alpha|\mu}\Gamma^\nu_{\ |\beta]\lambda}\big),
  \label{metrical_riemann}
\end{equation} 
and the Levi-Civita connection $\Gamma^\alpha_{\ \mu\nu}$ is of the form $\partial g$. As with GR, one cannot formally fit the whole standard model (SM) into the matter Lagrangian $L_{\text{m}}(\Phi;g)$. This is an elementary but occasionally overlooked limitation of metric theories: the matter fields $\Phi$ must be tensorial representations of $\mathrm{GL}(4,\mathbb{R})$, and are thus \emph{bosonic}. Note also that while $\phi$ and $\psi$ are historically termed \emph{galileons}, the covariantisation of the theory with respect to $g_{\mu\nu}$ breaks the Galilean shift symmetry. In exchange, \eqref{horndeski_full_lagrangian} acquires diffeomorphism invariance and (like GR) may be interpreted as a geometric $\mathbb{R}^{1,3}$ gauge theory.

\section{Tetrad theories}\label{tetrad_theories}
Various other geometric gauge theories have been proposed. Promotion of the proper, orthochronous Lorentz rotations to a local symmetry yields the Poincar\'e gauge theory (PGT) of $\mathbb{R}^{1,3}\rtimes\mathrm{SO}^+(1,3)$. The geometric interpretation of PGT replaces $\mathcal{  M}$ with a spacetime of Riemann--Cartan type in order to accommodate \emph{torsion}. The modern picture~\cite{blagojevic2002gravitation,1998RSPTA.356..487L,2016JMP....57i2505L} is perhaps more commensurate with particle physics in assuming a \emph{flat} metric $\gamma_{\mu\nu}$ on Minkowski spacetime $\check{\mathcal{M}}$. Translations are gauged by the field $\smash{h_a^{\ \mu}}$ and its inverse $b^a_{\ \mu}$, where $h_a^{\ \mu}b^a_{\ \nu}\equiv \delta^\mu_\nu$ and $\smash{h_a^{\ \mu}b^c_{\ \mu}\equiv \delta^c_a}$. The Roman indices refer to an anholonomic, Lorentzian basis. Lorentz rotations are gauged by the field $\smash{A^{ab}_{\ \ \mu}}$. The fields $\smash{h_a^{\ \mu}}$ and $\smash{A^{ab}_{\ \ \mu}}$ can be \emph{geometrically} interpreted as the tetrad and spin connection. They provide two field strengths 
\begin{subequations}
  \begin{align}
    \mathcal{  R}^{ab}_{\ \ \ cd}&\equiv 2h_c^{\ \mu}h_d^{\ \nu}\big(\partial^{\vphantom{ab}}_{[\mu} A^{ab}_{\ \ \ \nu]}+A^a_{\ e[\mu}A^{eb}_{\ \ \ \nu]}\big),\label{riemanndef}\\
    \mathcal{  T}^{a}_{\ \ bc}&\equiv 2h_b^{\ \mu}h_c^{\ \nu}\big(\partial^{\vphantom{d}}_{[\mu}b^{a\vphantom{d}}_{\ \nu]}+A^{a\vphantom{d}}_{\ d[\mu}b^d_{\ \nu]}\big),\label{torsiondef}%
  \end{align}
\end{subequations}
which are referred to as Riemann and torsion tensors, but which \emph{confer no geometry} to $\smash{\check{\mathcal{  M}}}$. The Ricci tensor $\mathcal{  R}^a_{\ b}\equiv\mathcal{  R}^{ac}_{\ \ \ bc}$, Ricci scalar $\mathcal{  R}\equiv\mathcal{  R}^a_{\ a}$ and torsion contraction $\mathcal{  T}_{a}\equiv\mathcal{  T}^b_{\ \ ab}$ are then used to construct the most general total Lagrangian up to quadratic order in the field strengths and invariant under parity inversions
  \begin{equation}
    \begin{aligned}
      L_{\text{T}}= &\alpha_1\mathcal{  R}^2+\mathcal{  R}_{ab}\big(\alpha_2\mathcal{  R}^{ab}+\alpha_3\mathcal{  R}^{ba}\big)\\
      &+\mathcal{  R}_{abcd}\big(\alpha_4\mathcal{  R}^{abcd}+\alpha_5\mathcal{  R}^{acbd}+\alpha_6\mathcal{  R}^{cdab}\big)\\
      &+{m_{\text{p}}}^2\big[\mathcal{  T}_{abc}\big(\beta_1\mathcal{  T}^{abc}+\beta_2\mathcal{  T}^{bac}\big)+\beta_3\mathcal{  T}_a\mathcal{  T}^a\big]\\
	&-\tfrac{1}{2}{m_{\text{p}}}^2\alpha_0\mathcal{  R}+L_{\text{m}}(\Phi,\Psi;h,A).
    \label{pgt_full_lagrangian}
    \end{aligned}
  \end{equation}
This general theory is termed PGT\textsuperscript{q,+}, and is parameterised by ten dimensionless coupling constants. Note that the remaining \emph{fermionic} fields $\Psi$ of the SM are now permitted in $L_{\text{m}}(\Phi,\Psi;h,A)$ as representations of $\mathrm{SL}(2,\mathbb{C})$, which universally covers $\mathrm{SO}^+(1,3)$. The Maxwell-like terms in \eqref{pgt_full_lagrangian} are motivated by analogy to the Yang-Mills structure of the SM: since \cref{riemanndef,torsiondef} are at lower order than \eqref{metrical_riemann}, maximally second-order field equations are guaranteed by construction.

\section{Scale-invariance}\label{scale_invariance}
Pushing the SM analogy further, one considers scale-invariance. This pertains to local conformal (or Weyl) transformations
\begin{subequations}
  \begin{gather}
    g_{\mu\nu}\mapsto \Omega^2g_{\mu\nu}, \quad \phi\mapsto \Omega^{-1}\phi, \quad \psi\mapsto \Omega^{-1}\psi,\label{metrical_conformal}\\
    b^a_{\ \mu}\mapsto \Omega b^a_{\ \mu}, \quad A^{ab}_{\ \ \mu}\mapsto A^{ab}_{\ \ \mu}.\label{weyl}
  \end{gather}
\end{subequations}
The Lagrangia \eqref{horndeski_full_lagrangian} and \eqref{pgt_full_lagrangian} are scale-invariant if they transform with weight $-4$, which cancels with the measure $\sqrt{|g|}$, or $h^{-1}\equiv\det b^a_{\ \mu}$. A scale-invariant PGT\textsuperscript{q,+} has $\alpha_0=\beta_1=\beta_2=\beta_3=0$, which eliminates the explicit mass scale $m_{\text{p}}$. By convention, $\phi$ and $\psi$ have weight $-1$~\cite{2014PhRvD..89f5009P} and $A^{ab}_{\ \ \mu}$ has weight $0$~\cite{2016JMP....57i2505L}. As a slight aside, an inhomogeneously rescaling $\smash{A^{ab}_{\ \ \mu}}$ was recently used in an \emph{extension} of \emph{Weyl} gauge theory (eWGT)~\cite{2016JMP....57i2505L}. Quite unlike PGT, eWGT is scale-invariant \emph{by construction}. However, when expressed in terms of scale-invariant variables~\cite{2019arXiv191101696S,1973PThPh..50.2080U,1975PThPh..53..565U}, the quadratic, parity-preserving version (eWGT\textsuperscript{q,+}) was shown to be dynamically equivalent to PGT\textsuperscript{q,+} under the SCP~\cite{2020arXiv200302690B}. At this level, PGT\textsuperscript{q,+} and eWGT\textsuperscript{q,+} differ only through a scale-dependent interpretation of the coupling constants. We will briefly return to eWGT\textsuperscript{q,+} in closing.

\section{The full metrical analogue}\label{the_full_metrical_analogue}
We will now construct an instance of \eqref{horndeski_full_lagrangian} which mimics \eqref{pgt_full_lagrangian} under the spatially-flat SCP. Adopting dimensionful Cartesian coordinates on $\mathcal{  M}$, the flat FLRW metric has interval
\begin{equation}
  \mathrm{d}s^2=\mathrm{d}t^2-a^2\mathrm{d}\mathbf{x}^2.
  \label{flat_metric}
\end{equation}
The dimensionless scale factor $a$ provides the Hubble number $H=\partial_ta/a$. Under conformal transformations of the form \eqref{metrical_conformal}, the form of \eqref{flat_metric} is always preserved by implicit combination with the diffeomorphism 
\begin{equation}
  \mathrm{d}t\mapsto \Omega^{-1} \mathrm{d}t, \quad H\mapsto \Omega^{-1}( H-\partial_t\Omega).
  \label{time_repar}
\end{equation}
Analogous Cartesian coordinates $\gamma_{\mu\nu}=\eta_{\mu\nu}$, assumed to transform according to \eqref{time_repar} under Weyl rescalings of the form \eqref{weyl}, then allow us to equate component values $g^{\mu\nu}\aneq\eta^{ab}h_a^{\ \mu}h_b^{\ \nu}$ and $g^{\phantom{\mu}}_{\mu\nu}\aneq\eta^{\phantom{\mu}}_{ab}b^a_{\ \mu}b^b_{\ \nu}$. Our `analogue equality' flags the notational abuse of incompatible tangent spaces. The torsion tensor on $\check{\mathcal{  M}}$ is restricted by the SCP to the scalar $U$ and pseudoscalar $Q$, which are the $0^+$ and $0^-$ sectors~\cite{1979PhLA...75...27T,2006gr.qc.....1089B,2008CQGra..25x5016B}
\begin{equation}
  \mathcal{  T}^a_{\ \ bc}=\delta^d_0\big( \tfrac{2}{3}U\delta^a_{[c}\eta^{\vphantom{a}}_{b]d}-Q\varepsilon^a_{\ dbc} \big).
  \label{torsion_definition}
\end{equation}
These are homogeneous cosmological fields in the same sense as $\phi$ and $\psi$, inviting the analogue of torsion on $\mathcal{  M}$
\begin{equation}
  \phi\aneq\tfrac{2}{3}U-2H, \quad \psi\aneq Q.
  \label{scalar_definition}
\end{equation}
Related constructions are used in~\cite{2020arXiv200302690B,2019arXiv190604340Z,lasenby-doran-heineke-2005} for algebraic convenience. In our case we see that \eqref{scalar_definition} corrects the inhomogeneous rescaling of $\mathcal{  T}^a_{\ \ bc}$, endowing the galileons with a weight of $-1$. Thus, all relations in \eqref{metrical_conformal} are reconciled with those in \eqref{weyl}. Finally, we tacitly convert matter fermions into bosons so as to preserve the stress-energy tensor $\smash{2\big(\tfrac{\delta}{\delta g}\big)\vphantom{g}_{\mu\nu}\big[\sqrt{|g|}L_{\text{m}}(\Phi;g)\big]}\aneq \eta^{\phantom{(a}}_{ab}b^a_{(\mu}\big(\tfrac{\delta}{\delta h}\big)\vphantom{h}^a_{\ \nu)}\big[h^{-1}L_{\text{m}}(\Phi,\Psi;h,A)\big]$, see e.g.~\cite{2006PrPNP..56..340M}. The spin tensor $\big(\tfrac{\delta}{\delta A}\big)\vphantom{A}_{ab}^{\ \ \mu}\big[h^{-1}L_{\text{m}}(\Phi,\Psi;h,A)\big]$ is \emph{neglected}. 

  At this point we are ready to derive the specific $\smash{G_2^{\phantom{\psi}}}$ and $\smash{G_4^{\phantom{\psi}}}$ which facilitate \eqref{pgt_full_lagrangian}. Throughout the PGT\textsuperscript{q,+} equations, the nine Maxwell-like couplings appear exclusively in \emph{five} linear combinations under the SCP
\begin{equation}
  \begin{gathered}
    \sigma_1\equiv\tfrac{3}{2}\alpha_1+\tfrac{1}{4}\alpha_2+\tfrac{1}{4}\alpha_3+\tfrac{1}{4}\alpha_5-\tfrac{1}{2}\alpha_6,\\
    \sigma_2\equiv\tfrac{3}{2}\alpha_1+\tfrac{1}{2}\alpha_2+\tfrac{1}{2}\alpha_3+\tfrac{3}{2}\alpha_4-\tfrac{1}{4}\alpha_5+\tfrac{1}{2}\alpha_6,\\
    \sigma_3\equiv\tfrac{3}{2}\alpha_1+\tfrac{1}{2}\alpha_2+\tfrac{1}{2}\alpha_3+\tfrac{1}{2}\alpha_4+\tfrac{1}{4}\alpha_5+\tfrac{1}{2}\alpha_6,\\
    \upsilon_1\equiv-2\beta_1+2\beta_2, \quad \upsilon_2\equiv2\beta_1+\beta_2+3\beta_3.
  \end{gathered}
  \label{cosmological_coordinates}
\end{equation}
These \emph{physical} couplings are insensitive to e.g. a Gauss--Bonnet variation $4\delta\alpha_1=-\delta\alpha_3=4\delta\alpha_6$, which is topological in $D\leq 4$. An application of the minisuperspace method is sufficient to obtain the required mapping from \eqref{pgt_full_lagrangian} to \eqref{horndeski_full_lagrangian}. For our previous treatments of the minisuperspace formulation of PGT\textsuperscript{q,+}, see \cite{2020arXiv200302690B,lasenby-doran-heineke-2005}. 
We use an ADM-like interval $\mathrm{d}s^2=u^2(\mathrm{d}t^2-v^2\mathrm{d}\mathbf{x}^2)$, where the flat FLRW interval in \eqref{flat_metric} is recovered by taking $u\mapsto 1$ and $v\mapsto a$. The analogue defined in \eqref{scalar_definition} corresponds to the following choices of gauge, in a further abuse of notation which assumes the holonomic and anholonomic bases to be aligned
\begin{subequations}
\begin{align}
  b^a_{\ \mu}& \aneq u\big( v(\delta^a_\mu-\delta^a_0\eta^{\vphantom{a}}_{0\mu}) +\delta^a_0\eta^{\vphantom{a}}_{0\mu} \big),\label{msso}\\
  A^{ab}_{\ \ \mu}& \aneq uv\delta^d_0\big( \phi\delta_\mu^{[b}\delta_d^{a]}-\tfrac{1}{2}\psi\varepsilon_{\mu d}^{\ \ \ ab} \big).
  \label{mss}
\end{align}%
\end{subequations}
The gauge fields in \eqref{mss} and \eqref{msso} are then substituted into \eqref{riemanndef} and \eqref{torsiondef}, and then into \eqref{pgt_full_lagrangian}. The Maxwell-like couplings defined in \eqref{cosmological_coordinates}, along with a minimal addition of surface terms (including the Gauss-Bonnet derivative) then reduce this to
\begin{widetext}
\begin{equation}
  \begin{aligned}
    L_{\text{T}}&\aneq \big(\tfrac{1}{2}{m_{\text{p}}}^2\upsilon_2+\sigma_3\phi^2+\tfrac{1}{2}(\sigma_3-\sigma_2)\psi^2 \big)\big[6v^3(\partial_tu)^2+12uv^2\partial_tu\partial_tv+6u^2v(\partial_tv)^2\big]%
			\\&\phantom{=}\ %
			+12\sigma_3\big[uv^3\phi\partial_tu+\tfrac{1}{2}u^2v^3\partial_t\phi+u^2v^2\phi\partial_tv\big]\partial_t\phi+6(\sigma_3-\sigma_2)\big[uv^3\psi\partial_tu+\tfrac{1}{2}u^2v^3\partial_t\psi+u^2v^2\psi\partial_tv\big]\partial_t\psi    %
			\\&\phantom{=}\ %
			+4\sigma_1\big(\psi^2-\phi^2\big)\big[\tfrac{3}{2}u^2v^3\phi\partial_tu+\tfrac{3}{2}u^3v^2\phi\partial_tv+\tfrac{3}{2}u^3v^3\partial_t\phi\big]+3{m_{\text{p}}}^2(\alpha_0+\upsilon_2)\big[u^2v^3\phi\partial_tu+u^3v^2\phi\partial_tv\big]    %
			\\&\phantom{=}\ %
			+\tfrac{3}{4}u^4v^3\big[ 2\sigma_3\phi^4-4\sigma_2\phi^2\psi^2+2\sigma_3\psi^4+{m_{\text{p}}}^2(\alpha_0+\upsilon_2)\phi^2-{m_{\text{p}}}^2(\alpha_0-4\upsilon_1)\psi^2  \big]+L_{\text{m}}(\Phi,\Psi;u,v,\phi,\psi).%
  \end{aligned}
  \label{mss_lagrangian}
\end{equation}%
\end{widetext}

A na\"ive ansatz restricts to \emph{polynomial} $G$-functions, but inspection of \eqref{mss_lagrangian} reveals that this is only viable up to surface terms if $\alpha_0+\upsilon_2=\sigma_1=0$.
These constraints eliminate terms of first order in $\partial_t\phi$ and $H$ from the penultimate line of \eqref{mss_lagrangian}, and so from the E.o.Ms. Such terms are non-canonical, but can be included (and the constraints removed) by extending \eqref{horndeski_full_lagrangian} to ${L_{\text{T}}\mapsto L_{\text{T}}+\Delta L_{\text{T}}}$, where
  \begin{equation}
    \begin{aligned}
      \Delta L_{\text{T}}&=\big[G_6^{\phi}(\phi,\psi)\partial_\mu\phi+G_6^{\psi}(\phi,\psi)\partial_\mu\psi\big] B^\mu\\
      &\phantom{=}+m_{\text{p}}\big({m_{\text{p}}}^2-B_\mu B^\mu\big)\chi.
      \label{hand}%
    \end{aligned}
  \end{equation}
The neutral vector $B^\mu$ and scalar $\chi$ may be thought of as gravitational \emph{spurions}: they constrain the theory by singling out a preferred timelike vector under the SCP without breaking general covariance in the action~\cite{2004EPJC...34..447H}. The spurions are generally non-dynamical and are integrated out directly such that \eqref{hand} merely renormalises $G_2$. Writing out the final $G$-functions explicitly, the full MA of \eqref{pgt_full_lagrangian} is
  \begin{subequations}
    \begin{align}
      \begin{split}
	L_{\text{T}}&=\big[ \tfrac{1}{2}{m_{\text{p}}}^2\upsilon_2+\sigma_3\phi^2+\tfrac{1}{2}(\sigma_3-\sigma_2)\psi^2 \big]R\\
	&\phantom{=}+12\big[\sigma_3X^{\phi\phi}+\tfrac{1}{2}(\sigma_3-\sigma_2)X^{\psi\psi}\big]+\sqrt{\vphantom{X^a_a}\smash[b]{|J_\mu J^\mu|}}\\
	&\phantom{=}+\tfrac{3}{4}{m_{\text{p}}}^2\big[(\alpha_0+\upsilon_2)\phi^2-(\alpha_0-4\upsilon_1)\psi^2\big]\\
	&\phantom{=}+\tfrac{3}{2}\big(\sigma_3\phi^4-2\sigma_2\phi^2\psi^2+\sigma_3\psi^4\big)+L_{\text{m}}(\Phi;g),
      \end{split}\\
      J_\mu&\equiv4\sigma_1\psi^3\partial_\mu(\phi/\psi)-{m_{\text{p}}}^2(\alpha_0+\upsilon_2)\partial_\mu\phi.
    \end{align}
    \label{final}%
  \end{subequations}
Further surface terms distinguish the minisuperspace Lagrangian of \eqref{final} from \eqref{mss_lagrangian}, and a straightforward calculation confirms that the E.o.Ms coincide with those of PGT\textsuperscript{q,+} under the spatially-flat SCP.
\section{First impressions}
Noting in what follows that $\sqrt{|J_\mu J^\mu|}$ carries an implicit factor of $\text{sgn}(J^0)$ for continuity~\cite{2014PhRvD..89h3521T}, a straightforward calculation confirms that \eqref{final} and \eqref{pgt_full_lagrangian} are dynamically coincident under the spatially-flat SCP. In this paper we will not consider inhomogeneous applications, e.g. to acoustic stability. Various features of the MA are already apparent at the Lagrangian level. Since $\smash{G_4^{\phantom{\psi}}}$ is not constant, $\phi$ and $\psi$ are non-minimally coupled to $R$, thus the MA has been unwittingly but naturally constructed in the \emph{Jordan} conformal frame (JF). It will prove convenient later to transform to the \emph{Einstein} frame (EF), but since the EF derives its meaning from the artificial context of the MA, we cannot take it to be physical. Equivalently, to work at the usual level of the PGT\textsuperscript{q,+} equations is to work in the JF of the MA \emph{and know no better}. While counter-intuitive, we find this picture to be unavoidable~\cite{1999astro.ph.10176F}. 
A scale-invariant PGT\textsuperscript{q,+} sets $\alpha_0=\upsilon_1=\upsilon_2=0$, reducing the MA to a \emph{manifestly} conformal field theory~\cite{2014PhRvD..89f5009P}. In our minimal formulation, this would restrict to a pure radiation cosmology (see e.g.~\cite{lasenby-doran-heineke-2005}), but we note that various Higgs-like scale symmetry-breaking extensions to the gauge theory have been proposed~\cite{1973RSPSA.333..403D,1977PThPh..58.1627O,1982PhLB..109..435S}.

\section{Application to established theories}\label{application_to_established_theories}
Before addressing the novel theories, we will analyse some `conventional' PGT\textsuperscript{q,+}s with non-dynamical $\smash{A^{ab}_{\ \ \mu}}$. Consider the representative two-parameter theory 
\begin{equation}
  L_{\text{T}}=-\tfrac{1}{2}{m_{\text{p}}}^2\alpha_0\mathcal{  R}+\tfrac{1}{2}{m_{\text{p}}}^2\beta\mathbb{T}+L_{\text{m}}(\Phi,\Psi;h,A),
  \label{spectrum}
\end{equation}
i.e. a linear combination of $\mathcal{  R}$ and the \emph{teleparallel} term $\mathbb{T}\equiv\tfrac{1}{4}\mathcal{  T}_{abc}\mathcal{  T}^{abc}+\tfrac{1}{2}\mathcal{  T}_{abc}\mathcal{  T}^{bac}-\mathcal{  T}_a\mathcal{  T}^a$, with the MA
\begin{equation}
  \begin{aligned}
    L_{\text{T}}&=-\tfrac{1}{2}{m_{\text{p}}}^2\beta R+{m_{\text{p}}}^2(\beta-\alpha_0)\big[\smash{\sqrt{\vphantom{X_x}\smash{2|X^{\phi\phi}|}}}\\
      &\ \phantom{=}-\tfrac{3}{4}\phi^2+\tfrac{3}{4}\psi^2\big]+L_{\text{m}}(\Phi;g).\\
  \end{aligned}
  \label{spectrumma}
\end{equation}
We see that the MA is a linear combination of $R$, a quadratic \emph{Cuscuton} $\phi$ with equation of motion $\phi=-2H$, and a non-dynamical mass which sets $\psi=0$. By \eqref{scalar_definition} we will have $U=Q=0$. As a general principle, the \emph{Cuscuton} is a non-dynamical constraint field, and preserves the form of the usual Friedmann equations of GR that follow from $R$. This can be seen by substituting $\phi$ into the $g_{\mu\nu}$ equation of \eqref{spectrumma}~\cite{2007PhRvD..75l3509A}. ECKS theory is equivalent to GR when the spin tensor vanishes, and is defined by $\alpha_0=1$ and $\beta=0$ in \eqref{spectrum}~\cite{2006gr.qc.....6062T}. Remarkably, this eliminates $R$ from \eqref{spectrumma} entirely, so that $\mathcal{  R}$ is represented purely by the \emph{Cuscuton}. If $\beta\neq 0$, the admixture of $\mathbb{T}$ in \eqref{spectrum} leads to $R$--\emph{Cuscuton} contributions in \eqref{spectrumma} which \emph{exactly cancel} in the $g_{\mu\nu}$ equation. However, true teleparallelism, with $\beta=1$ and $\alpha_0=0$ is also dynamically equivalent to GR if PGT curvature (as defined in \eqref{riemanndef}) \emph{vanishes} identically~\cite{2018arXiv180106929G,2019PhRvD.100f4018B,blagojevic2002gravitation}. The constraint $\smash{\mathcal{  R}^{ab}_{\ \ \ cd}\equiv 0}$ is properly imposed via Lagrange multiplier fields~\cite{blagojevic2002gravitation}, but in practice this just restricts $\smash{A^{ab}_{\ \ \mu}}$ to a pure gauge (the \emph{Weitzenb\"ock connection}) and fixes $\phi\equiv\psi\equiv 0$. By \eqref{scalar_definition} we will then have $Q\equiv 0$ and $U\equiv 3H$. Since the \emph{Cuscuton} is now eliminated, $\mathbb{T}$ is represented purely by $R$, and the expected equivalence to GR is immediate.

\begin{figure}[t]
  \includegraphics[width=\linewidth]{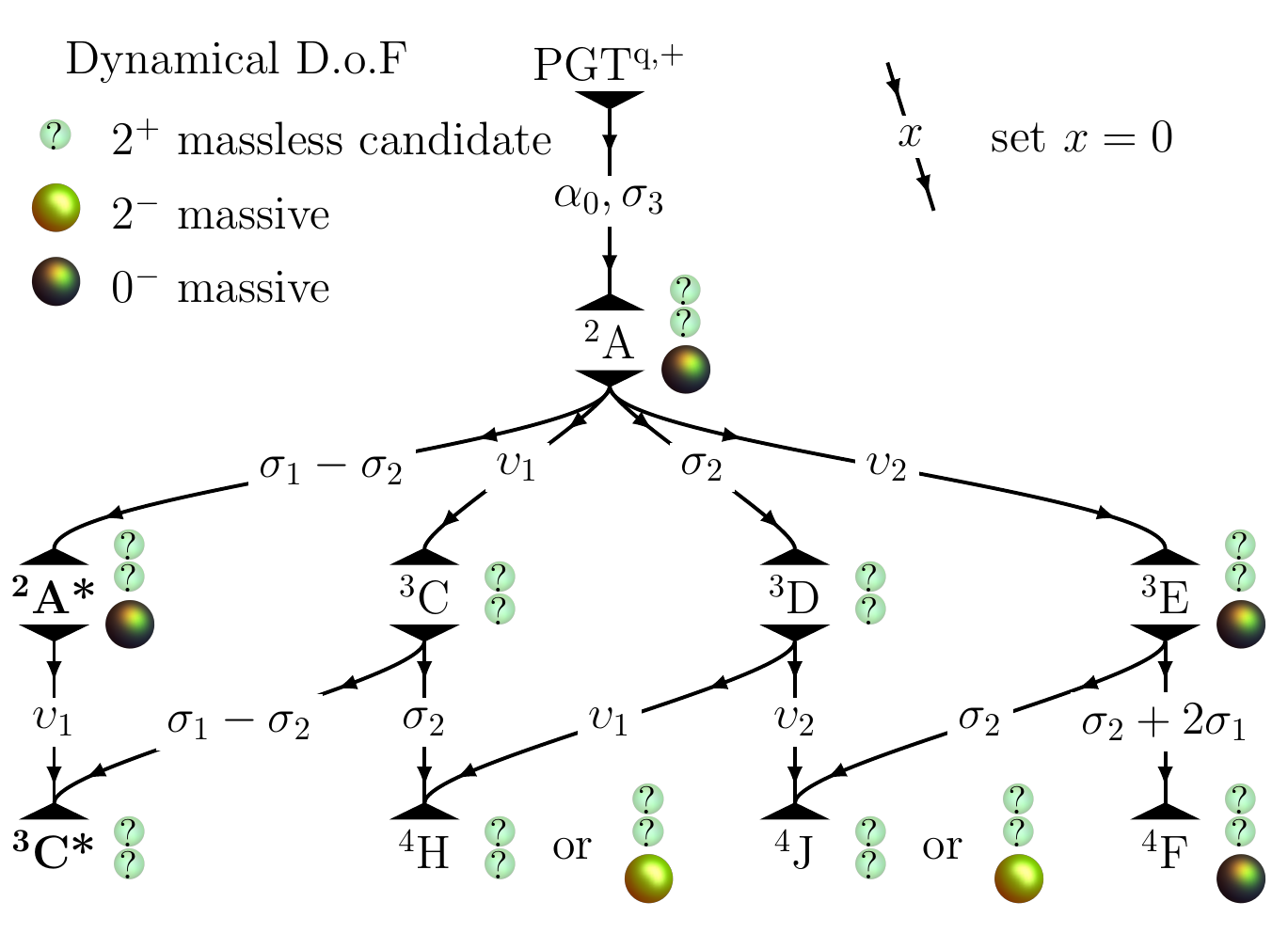}
  \caption{\label{map}Cosmologies and associated particle content of the novel theories (see \cite{2020PhRvD.101f4038L} and literature comparisons in \cite{2019PhRvD..99f4001L,2020arXiv200302690B}). In the weak, free-field limit, certain cases of PGT\textsuperscript{q,+} are unitary and power-counting-renormalisable. These cases contain propagating irreps of $\mathrm{SO}(3)$, i.e. D.o.F of spin-parity $J^P$. For massless D.o.F, the propagator poles associated with any contributing $J^P$ sectors are degenerate at the origin of $p$-space. Since this leads to ambiguity, we restrict to cases which \emph{do not preclude} the two $2^+$ polarisations of the graviton (which should be unique~\cite{Porrati:2012rd}). The cases are grouped into cosmological \emph{classes}, of which we consider \cosmicclass{cnull} and \cosmicclass{null}.}%
\end{figure}%

\section{Application to novel theories}\label{application_to_novel_theories}
The graviton and roton propagators of a generic PGT\textsuperscript{q,+} may approach the UV as $p^{2N_h}$ and $p^{2N_A}$, where $N_h,N_A\leq 0$ are some integers, and the even powers are expected of bosons. In such a theory, a diagram may have $E_h$ external graviton and $E_A$ external roton lines. Also, there will be $V_{nm}$ vertices with $n$ graviton and $m$ roton valences, and whose coupling constant has some (low) mass dimension $C_{nm}$ supplied by the appearance of $m_{\text{p}}$ in \eqref{pgt_full_lagrangian}. By considering the perturbative structure of \eqref{pgt_full_lagrangian} and applying the usual topological identity that relates the number propagators, vertices and loops \cite{1995iqft.book.....P}, one eventually arrives at the following formula for the superficial divergence $D$ of the diagram 
\begin{align}
  \begin{split}
  D&=4-(2+N_h)E_h-(2+N_A)E_A\\
  &-\sum_{n,m}\left[C_{nm}-2n(2+N_h)-2m(1+N_A) \right]V_{nm}.
\end{split}
\label{superficial}
\end{align}
The strong PCR criteria $N_h=-2$ and $N_A=-1$ are then \emph{suggestive} of perturbative renormalisability. If these criteria are met, one can see from \eqref{superficial} that any diagram appearing at high enough loop order or with sufficiently many external lines should superficially converge. While such a diagram may still be divergent in practice, there is some hope that this divergence may result from the incorporation of a finite number of primitively divergent diagrams. The novel cases in~\cite{2019PhRvD..99f4001L,2020PhRvD.101f4038L} are defined by linear constraints on the ten PGT\textsuperscript{q,+} parameters. These constraints structurally alter the saturated propagator, obtained by inverting the linearised, matter-free Lagrangian in \eqref{pgt_full_lagrangian}, so as to \emph{effectively} satisfy these criteria. As noted in \cref{introduction}, the criteria may be safely relaxed for modes which become non-propagating in the UV; for a full discussion of this matter the reader is referred to~\cite{2020PhRvD.101f4038L}.

The SCP groups the cases into \emph{classes}, some of which are shown in \cref{map}. The constraint $\alpha_0=0$ marks a complete break with ECKS theory: one is left only with quadratic invariants which have no EFT interpretation as loop corrections to the PGT Ricci scalar $\mathcal{  R}$. The further constraint $\sigma_3=0$ then triggers the $k$-screening mechanism, in which the physical spatial curvature $k\in\{\pm 1,0\}$ is \emph{eliminated} from the PGT\textsuperscript{q,+} equations: a hyperspherical, hyperbolic or simply flat choice of universe does not affect the background dynamics~\cite{2020arXiv200302690B}. The description of such classes as offered by the MA is thus \emph{not limited} by our earlier assumption of spatial flatness in \eqref{flat_metric}.

We consider \cosmicclass{cnull}, defined by the further constraint $\sigma_2=\sigma_1$ (note that \cosmicclass{null} will always be the special case $\upsilon_1=0$). We next set $\sigma_1<0$ (no ghost) and $\upsilon_1<0$ (no tachyon): these unitarity conditions are translated from~\cite{2020PhRvD.101f4038L}. They may also be read off from \eqref{final} near the vacuum ${R=\phi=\psi=0}$, once the defining constraints are imposed. We finally take a third condition $\upsilon_2<0$ by analogy to the Einstein-Hilbert Lagrangian, although this is not listed in \cite{2020PhRvD.101f4038L}. A conformal transformation $\Omega$ takes the MA of \cosmicclass{cnull} into the EF. Following the conventions of e.g. Brans--Dicke theory~\cite{2007MPLA...22..367B}, we then partly recanonicalise the MA through two new fields $\zeta(\phi,\psi)$ and $\xi(\psi)$
  \begin{subequations}
    \begin{align}
      \begin{split}
      L_{\text{T}}& = -\tfrac{1}{2}{m_{\text{p}}}^2R+X^{\xi\xi}+{m_{\text{p}}}^2\omega(\xi)^3\smash{\sqrt{\vphantom{X_x}\smash{|X^{\zeta\zeta}|}}}\\
    &\phantom{=}\ -V(\xi)+\tfrac{3}{4}{m_{\text{p}}}^2\omega(\xi)^4\zeta^2+L_{\text{m}}(\Phi,\xi;g),\label{as_lag_central}
  \end{split}\\
      V(\xi)&\equiv-\tfrac{4\upsilon_1}{3\sigma_1\upsilon_2}{m_{\text{p}}}^4\big(1+\tfrac{1}{8}\omega(\xi)^2\big)\big(1+\tfrac{1}{2}\omega(\xi)^2 \big),\label{Vdef}\\
      \omega(\xi)&\equiv\sqrt{\big|3\cosh\big(\sqrt{2/3} \ \xi/m_{\text{p}}\big)-5\big|}.\label{Wdef}
    \end{align}
    \label{as_lag}%
  \end{subequations}
  While \eqref{as_lag} is strictly valid for the range
  \begin{equation}
    1\leq4\sigma_1Q^2/\upsilon_2{m_{\text{p}}}^2<4,
    \label{validity}
  \end{equation}
we will use it to obtain physical results which are completely general, as may be confirmed directly from \eqref{final}. In fact, we will later see that the universe is expected to lie in this range for most of its history anyway. 
Noting that $\smash{\Omega^2=-\tfrac{4}{3\upsilon_2}\big(1+\tfrac{1}{8}\omega^2\big)}$, it seems natural in what follows to take $\upsilon_2=-4/3$, and this choice will be justified in stages. The `conformal shift' $\omega$ now measures the degree to which the physical JF has strayed from the EF, and so mediates any $\xi$--$\Phi$ coupling. Note that $\omega$ also weights the field $\zeta$, which is a quadratic \emph{Cuscuton}. The field $\xi$ is canonical, and in moving from \cosmicclass{null} to \cosmicclass{cnull} it acquires a potential $V$. Note that $V$ traces back to the mass of $\psi$, which in turn corresponds to the massive $0^-$ D.o.F in \cref{map}. By inspection, $V$ must act as a (quintessence) dark energy source, since $\upsilon_1/\sigma_1>0$.
In the final sections we will make the nature of this dark energy more concrete, using the $\zeta$ E.o.M as a heuristic
  \begin{equation}
    \omega^2\big(\sqrt{2} \partial_\xi \omega\partial_t\xi+\sqrt{2}\omega H-\omega^2\zeta \big)=0.
    \label{stallme}
  \end{equation}

  \section{Negative screened dark energy}\label{negative_screened_dark_energy}
By analogy to \eqref{spectrumma}, suppose that the \emph{Cuscuton} obeys $\zeta\propto H$, which was its `minimally-coupled' behaviour. This is possible if the last two terms in \eqref{stallme} cancel, whereupon the decay of $\xi$ \emph{stalls} above the natural vacuum of $V$ at constant conformal shift ${\omega=\sqrt{2}H/\zeta}$. This solution has the following utility if the physical JF matter Lagrangian contains only a bare cosmological constant ${L_{\text{m}}(\Phi;g)=-{m_{\text{p}}}^2\Lambda_{\text{b}}}$. Accelerated expansion is difficult to drive with ${\Lambda_{\text{b}}<0}$ in many gravitational theories. This can make them hard to reconcile with attractive, more fundamental theories~\cite{2012arXiv1205.3807H,2014JHEP...06..095M,2009PhRvD..80b3519B,2011arXiv1105.0078P}. If ${L_{\text{m}}(\Phi,\xi;g)=-{m_{\text{p}}}^2\Lambda_{\text{b}}\smash{\big(1+\tfrac{1}{8}\omega^2\big)\vphantom{\omega}^2}}$ and ${\omega=\sqrt{2}H/\zeta}$ are substituted into the remaining E.o.Ms of \eqref{as_lag}, one can straightforwardly solve for $\xi$ and $H$ in the EF. In the physical JF this gives $Q^2=2\Lambda_{\text{b}}/3\upsilon_1$, and ${H^2=\Lambda/3}$, where the effective cosmological constant is $\Lambda=\upsilon_1{m_{\text{p}}}^2/2\sigma_1$. Remarkably therefore, a \emph{negative} $\Lambda_{\text{b}}$ is required, yet \emph{screened} from the de Sitter expansion rate. 

\begin{figure}[t]
  \includegraphics[width=\linewidth]{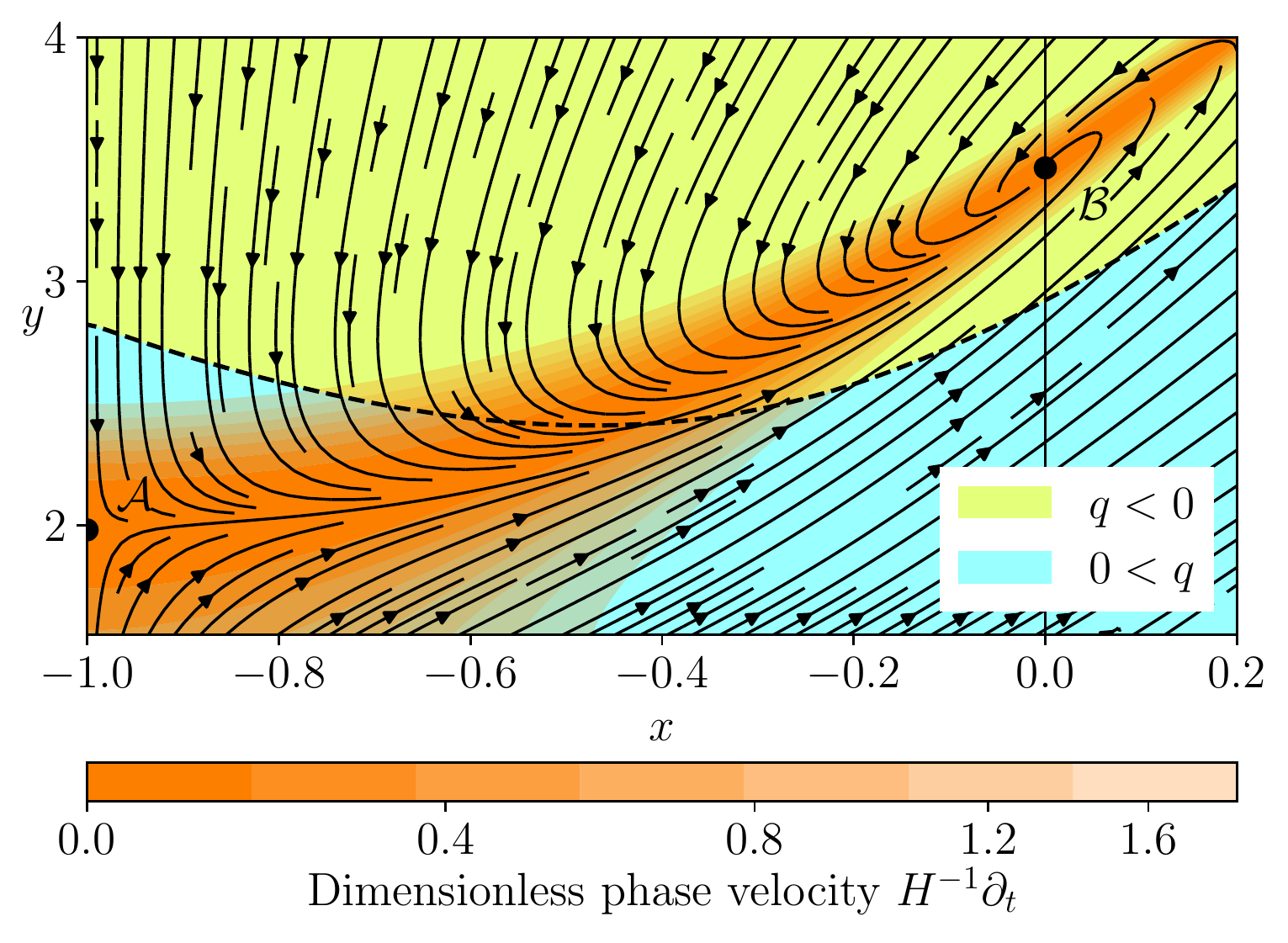}
  \caption{\label{phase}Partial phase portrait of \cosmicclass{cnull}, with \emph{negative} bare cosmological constant ${\Lambda_{\text{b}}=-0.48{m_{\text{p}}}^{2}}$. The saddle $\mathcal{A}$ deflects the universe towards the de Sitter attractor $\mathcal{B}$ in the inflationary region where it feels a \emph{positive} effective $\Lambda=0.1{m_{\text{p}}}^{2}$, all in the physical JF. The EF deceleration parameter is $1+q=-\partial_tH/H^2$. Hamiltonian coordinates $y$ and $x$ describe the $0^-$ torsional mode. Phase velocity reflects elapsing Hubble-times.}
\end{figure}

To verify the stability of the de Sitter solution, we employ another product of the MA: the powerful dynamical systems theory of scalar-tensor inflation~\cite{2014PhRvD..89h3521T,2001PhRvD..64h3510N}. We view $\xi$ as a canonical inflaton, whose `total potential' is ${V_{\text{T}}\equiv V+L_{\text{m}}(\Phi,\xi;g)}$. It is possible to encode all E.o.Ms as an autonomous, first order system in the dimensionless variables 
\begin{equation}
  x^2\equiv\frac{{m_{\text{p}}}^2(\partial_{t}\xi)^2}{6H^2}, \quad y^2\equiv \frac{V_{\text{T}}}{3{m_{\text{p}}}^2H^2},
  \label{<+label+>}
\end{equation}
which are the comoving Hamiltonian coordinates of $\xi$. In order to obtain this form, we further define intermediate dimensionless variables
\begin{equation}
  z^2\equiv\frac{{m_{\text{p}}}^2\omega^4\zeta^2}{4H^2}, \quad \lambda\equiv-\frac{m_{\text{p}}\partial_{\xi}V_{\text{T}}}{V_{\text{T}}}, \quad \mu\equiv\omega.
  \label{<+label+>}
\end{equation}
Note that $x$, $y$ and $\lambda$ are conventional parameters in the literature, while $\mu$ is defined for convenience and $z$ is somewhat analogous to the conventional matter parameter~\cite{2014PhRvD..89h3521T,2001PhRvD..64h3510N}. From \eqref{as_lag}, the $\xi$ equation (or alternatively the pressure--$g_{\mu\nu}$ equation) combined with the derivative of the $\zeta$ equation \eqref{stallme} can be expressed as a coupled first-order system in terms of these variables
\begin{widetext}
\begin{subequations}
    \begin{alignat}{2}
      \partial_\tau x&=\input{grad_x_refined.txt},\label{ffirst}\\
      \partial_\tau y&=\input{grad_y_refined.txt},\label{ssecond}
\end{alignat}
\label{systemeqs}%
\end{subequations}%
\end{widetext}
where the dimensionless (Hubble-normalised) time is $\mathrm{d}\tau=H\mathrm{d}t$. In order to obtain the autonomous system in $x$ and $y$ we must eliminate $\lambda$, $\mu$ and $z$ from \eqref{ffirst} and \eqref{ssecond}. Using \eqref{Vdef} and \eqref{Wdef}, it is possible to solve for $\lambda$ in terms of $\mu$
\begin{equation}
  \lambda=-\tfrac{\big[4\big(2\Lambda_{\text{b}}  + 5\tfrac{\upsilon_1}{\sigma_1}{m_{\text{p}}}^2\big)  + \big(\Lambda_{\text{b}}  + 4\tfrac{\upsilon_1}{\sigma_1}{m_{\text{p}}}^2 \big)\mu ^2\big]\sqrt{\big(2+\mu^2\big)}}{\big[8\big(\Lambda_{\text{b}}  + \tfrac{\upsilon_1}{\sigma_1}{m_{\text{p}}}^2\big)  + \big(\Lambda_{\text{b}}  + 4\tfrac{\upsilon_1}{\sigma_1}{m_{\text{p}}}^2 \big)\mu ^2\big]\sqrt{3\big(1+\tfrac{1}{8}\mu^2\big)}}.
  \label{lambda}%
\end{equation}%
Note that \eqref{lambda} explicitly incorporates both the bare cosmological constant $\Lambda_{\text{b}}$ and our central combination $\upsilon_1{m_{\text{p}}}^2/\sigma_1$. As emphasised above, these quantities should be considered on an equal footing. Next, the $\zeta$ equation reduces to a quartic in $\mu$
\begin{equation}
  \big(x^2-1\big)\mu^4+2\sqrt{2}z\mu^3+2\big(5x^2-z^2\big)\mu^2+16x^2=0.
  \label{quartic}%
\end{equation}%
Finally, $z$ is solved for $x$ and $y$ by the density--$g_{\mu\nu}$ equation
\begin{equation}
  x^2+y^2-z^2=0,
  \label{expell}%
\end{equation}%
revealing that the physical portions of the phase space are \emph{expelled} from the unit disc. If $z$ were a `conventional' matter parameter (i.e. proportional to a density which is obedient to the weak energy condition), the phase space would be \emph{confined} to the unit disc. This more holistic picture, in which all critical points $\partial_\tau x=\partial_\tau y=0$ are visible, may be reached by taking a simple M\"obius transform of the phase space. The quartic roots of \eqref{quartic} cause the fully autonomous system to be highly unwieldy. This is a natural consequence of explicitly encoding the \emph{Cuscuton} constraint in the \cosmicclass{cnull} and \cosmicclass{null} theories, rather than a generic limitation of the MA in \eqref{final}. Returning at last to the question of stability, the de Sitter solution outlined above is then found to be a stable critical point in this system, as illustrated in \cref{phase}. 

While $\zeta\propto H$ may describe our late universe if $\Lambda_\text{b}<0$, it is not self-consistent in a matter-dominated epoch. Therefore, we will next consider a family of solutions which naturally describe the whole expansion history.

\section{Generally viable dark energy}\label{generally_viable_dark_energy}
The `generally viable' solution to \eqref{stallme} occurs at vanishing conformal shift ${\omega=0}$, where the EF and physical JF \emph{coincide}. We previously termed this the \emph{correspondence solution} (CS)~\cite{2020arXiv200302690B}.
The CS of \cosmicclass{null} reduces \eqref{as_lag_central} to GR by inspection; \cosmicclass{cnull} differs from this through the constant stalled potential $V$. The stalled $\xi$ fixes ${Q^2=-{m_{\text{p}}}^2/3\sigma_1}$. If the universe is reasonably assumed to follow the CS closely, then $Q$ should not stray too far from this critical value, which is fortunately the lower bound in \eqref{validity}. The equivalence of conformal frames is guaranteed by our earlier condition ${\upsilon_2=-4/3}$. Broadly speaking, this has the same effect as fixing Einstein's $\kappa$ in GR. 

The stability of the CS should be verified for all matter in \textLambda CDM including the conventional $\Lambda_{\text{b}}\geq 0$, but the earlier dynamical systems approach is impractical in this case. Such matter may be characterised by linear equations of state (E.o.S) $\rho=wP$, diluting away as ${\rho\propto a^{-3(1+w)}}$. For any dominant matter, a straightforward perturbation around the CS is equivalent to adding an effective fluid ${\rho\mapsto\rho+\rho_{\text{eff}}}$ to GR. The effective E.o.S parameter \emph{tracks} the dominant $w$ according to
\begin{equation}
  w_{\text{eff}}(w)\equiv\tfrac{1}{2} (w+1)-\tfrac{1}{6}\sqrt{9w^2+3}, \quad -1\leq w\leq \tfrac{1}{3},
  \label{eff}
\end{equation}
The $\rho_{\text{eff}}$ becomes increasingly sub-dominant (and the CS is stable) when ${w_{\text{eff}}(w)>w}$; the only exception is co-dominant dark radiation, since ${w_{\text{eff}}(1/3)=1/3}$. The possible utility of this dark radiation in shrinking the sound horizon at recombination (and raising the early-universe inference of $\mathsf{h}$) is discussed in~\cite{2020arXiv200302690B}. Note that the \emph{effective} fluid need not satisfy the weak energy condition by itself. This strengthens the justification of \eqref{validity}, since a value of $Q$ below the lower bound would manifest as ${\rho_{\text{eff}}<0}$, i.e. a negative dark radiation fraction which would exacerbate the Hubble tension.
Finally, the stalled $V$ readily gives an effective ${\Lambda=\Lambda_{\text{b}}+\upsilon_1{m_{\text{p}}}^2/\sigma_1}$.

\section{Conclusions}\label{discussion}
We constructed in \eqref{final} a non-canonical bi-scalar-tensor theory, the \emph{metrical analogue} (MA) which lays bare the rich IR background cosmology of PGT\textsuperscript{q,+}.
It is natural that the theory explicitly includes only the cosmological $0^+$ and $0^-$ torsion sectors, rather than all $20$ D.o.Fs native to PGT. As a consequence, portions of both the IR and UV are necessarily lost. In particular, it is evident that no parameter constraint may be applied to the MA itself to render it perturbatively renormalisable. This follows since the MA is an explicit extension of GR by scalar D.o.Fs, and lacks any of the expected quadratic curvature invariants. However, we see no reason why this should affect the anticipated renormalisability of the underlying PGT\textsuperscript{q,+}. Rather, it is interesting to consider how the quadratic and linear invariants of PGT\textsuperscript{q,+} are allocated to the linear invariant of the MA. Tellingly, it is teleparallelism and the other quadratic theories which inherit the Einstein--Hilbert Lagrangian, while ECKS theory is relegated to a \emph{Cuscuton}. We verified that the Friedmann equations are recovered in both cases. This illustrates, in the context of our introductory discussion, the naturalness of \emph{quadratic} PGT Lagrangia.

Our analysis in this paper of the MA phenomenology was not intended to be exhaustive. Particularly, our approach invites inflationary applications in the early universe, and extension to Weyssenhoff fluids through a non-minimal $\psi$-coupling to modified matter sources~\cite{2020arXiv200406058I}. A principle observation is that PGT\textsuperscript{q,+}, when expressed in scalar-tensor form, contains a non-canonical term which may often be interpreted as a \emph{Cuscuton} field. While this interpretation offers theoretical support to the \emph{Cuscuton}, we note that it is not unique. For instance, it is evident from \eqref{hand} that by alternatively integrating out a galileon the MA would contain a neutral vector. Specifically, the physics is basically equivalent (as is the \emph{Cuscuton} itself) to the cosmological model of \emph{Lorentz-violating vector fields} \cite{2004PhRvD..70l3525C}.

In this paper we focussed on late-universe dark energy in recently proposed, superficially healthy cases of PGT\textsuperscript{q,+}.
The proposed emergent $\Lambda=\Lambda_{\text{b}}+\upsilon_1{m_{\text{p}}}^2/\sigma_1$ still does not address the `strong' cosmological constant problem~\cite{2012CRPhy..13..566M,2011arXiv1105.6296K}. Let us assume a `non-gravitating vacuum' $\Lambda_{\text{b}}=0$ \cite{1996PhRvD..53.7020G,2012CRPhy..13..566M,2019NuPhB.94614694E}. CMB-inference fixes ${\Lambda=\SI[separate-uncertainty=true,multi-part-units=single]{7.15(19)e-121}{\ccs}}$~\cite{2018arXiv180706209P}, with some (slight) shift expected from any dark radiation we may choose to add~\cite{2018JCAP...09..025M,2019JCAP...10..029S}. The requisite $\upsilon_1/\sigma_1\sim\SI{e-121}{\nothing}$ then reveals an apparent hierarchy. 
We tentatively observe that the hierarchy appears less severe in the scale-invariant eWGT counterpart, since the $\sim\SI{4.1}{\giga\parsec}$ Hubble horizon endows specific physical eWGT\textsuperscript{q,+} couplings with a natural length scale~\cite{2016JMP....57i2505L}. This builds the case for a future extension of the systematic analysis in~\cite{2019PhRvD..99f4001L,2020PhRvD.101f4038L,2020arXiv200502228L} to eWGT\textsuperscript{q,+}, whose propagator is currently unexplored.

In a conservative summary, the \cosmicclass{cnull} theory not only \emph{matches} the GR background, but can provide dark radiation and (hierarchical) dark energy. Unlike GR~\cite{Buoninfante:2016iuf}, the perturbative renormalisability of this unitary theory is not precluded by a simple power counting~\cite{2019PhRvD..99f4001L,2020PhRvD.101f4038L}; a nonlinear Hamiltonian analysis may offer further insight. The $0^-$ torsional mode must survive averaging over homogeneous comoving scales of ${\gtrsim\SI{300}{\per\littleh\mega\parsec}}$~\cite{2010MNRAS.405.2009Y,2018MNRAS.475L..20G}. This mode has yet to be constrained, even in an Earth-based laboratory~\cite{2010RPPh...73e6901N,1997PhLA..228..223L,2014IJMPD..2342004P}, and its strength is not separable here from the $\sigma_1$ or $\upsilon_1$ couplings. Indeed, the expansion history only determines $\upsilon_2$ and $\upsilon_1/\sigma_1$, which translate to the two freedoms in Lovelock's theorem.

\begin{acknowledgments}

We are grateful to Fernando Quevedo for essential comments following the original presentation of this work at the DAMTP GR Seminar Series on 14\textsuperscript{th} February 2020, Yun-Cherng Lin for assistance with the novel theories, David Tong and Miguel Zumalac\'arregui for helpful correspondence, and Amel Durakovic for valuable discussions and suggestions which improved the manuscript. WEVB is supported by the Science and Technology Facilities Council -- STFC under Grant ST/R504671/1, and WJH by a Gonville and Caius Research Fellowship.
\end{acknowledgments}

\bibliographystyle{apsrev4-1}
\bibliography{bibliography}

\end{document}

%% file: cosmic_class_names.tex
\newrobustcmd{\stall}[1]{%
\IfEqCase{#1}{%
  {1}{State I\xspace}%
  {2}{State II\xspace}%
}[\packageError{cosmicclass}{Unidentified Critical Case: #1}{}]%
}
\newrobustcmd{\numbercosmicclass}{14}%
\newrobustcmd{\cosmicclass}[1]{%
\IfEqCase{#1}{%
{1}{Class~\textsuperscript{3}E\xspace}%
{2}{Class~\textsuperscript{2}A\xspace}%
{8}{Class~\textsuperscript{4}H\xspace}%
{9}{Class~\textsuperscript{5}M\xspace}%
{10}{Class~\textsuperscript{5}M\xspace}%
{11}{Class~\textsuperscript{4}I\xspace}%
{12}{Class~\textsuperscript{4}J\xspace}%
{14}{Class~\textsuperscript{4}H\xspace}%
{13}{Class~\textsuperscript{4}K\xspace}%
{16}{Class~\textsuperscript{3}C\xspace}%
{15}{Class~\textsuperscript{3}D\xspace}%
{21}{Class~\textsuperscript{3}F\xspace}%
{20}{Class~\textsuperscript{3}F\xspace}%
{23}{Class~\textsuperscript{4}L\xspace}%
{22}{Class~\textsuperscript{3}F\xspace}%
{26}{Class~\textsuperscript{4}N\xspace}%
{27}{Class~\textsuperscript{3}E\xspace}%
{24}{Class~\textsuperscript{3}F\xspace}%
{25}{Class~\textsuperscript{3}F\xspace}%
{30}{Class~\textsuperscript{3}E\xspace}%
{31}{Class~\textsuperscript{3}G\xspace}%
{28}{Class~\textsuperscript{4}N\xspace}%
{29}{Class~\textsuperscript{2}A\xspace}%
{35}{Class~\textsuperscript{3}E\xspace}%
{34}{Class~\textsuperscript{3}G\xspace}%
{33}{Class~\textsuperscript{4}N\xspace}%
{32}{Class~\textsuperscript{4}N\xspace}%
{39}{Class~\textsuperscript{3}G\xspace}%
{38}{Class~\textsuperscript{3}G\xspace}%
{37}{Class~\textsuperscript{2}A\xspace}%
{36}{Class~\textsuperscript{2}B\xspace}%
{40}{Class~\textsuperscript{2}B\xspace}%
{41}{Class~\textsuperscript{2}A\xspace}%
{null}{Class~\textsuperscript{3}C*\xspace}%
{cnull}{Class~\textsuperscript{2}A*\xspace}%
}[\packageError{cosmicclass}{Unidentified Critical Case: #1}{}]%
}
\newrobustcmd{\criticalcase}[1]{%
\IfEqCase{#1}{%
{1}{Case 1\xspace}%
{2}{Case 2\xspace}%
{8}{Case 8\xspace}%
{9}{Case \textsuperscript{*1}9\xspace}%
{10}{Case \textsuperscript{*3}10\xspace}%
{11}{Case \textsuperscript{*4}11\xspace}%
{12}{Case 12\xspace}%
{14}{Case 14\xspace}%
{13}{Case \textsuperscript{*2}13\xspace}%
{16}{Case 16\xspace}%
{15}{Case 15\xspace}%
{21}{Case 21\xspace}%
{20}{Case 20\xspace}%
{23}{Case 23\xspace}%
{22}{Case 22\xspace}%
{26}{Case \textsuperscript{*6}26\xspace}%
{27}{Case 27\xspace}%
{24}{Case 24\xspace}%
{25}{Case \textsuperscript{*5}25\xspace}%
{30}{Case \textsuperscript{*7}30\xspace}%
{31}{Case \textsuperscript{*8}31\xspace}%
{28}{Case 28\xspace}%
{29}{Case 29\xspace}%
{35}{Case \textsuperscript{*9}35\xspace}%
{34}{Case 34\xspace}%
{33}{Case 33\xspace}%
{32}{Case 32\xspace}%
{39}{Case 39\xspace}%
{38}{Case 38\xspace}%
{37}{Case 37\xspace}%
{36}{Case \textsuperscript{*10}36\xspace}%
{40}{Case 40\xspace}%
{41}{Case 41\xspace}%
}[\packageError{criticalcase}{Unidentified Critical Case: #1}{}]%
}

%% file: grad_x_refined.txt
-\big[x\big(2\sqrt{3}\lambda\mu^3xy^2 + 4z^2\big(\big(\mu^4-8\big)z^2- 2\big(\mu^4-4\big)y^2 -8 \big)+ \sqrt{2}\mu\big(\sqrt{3}\lambda\mu^3xy^2z^2 + \mu^4(y - z)^2(y + z)^2\nonumber\\
& - 16(y - z)(y + z)\big(y^2 - z^2-1\big)+ 2\mu^2\big(y^2-1\big)\big(3y^2 - z^2\big)\big)\big)\big]/\big[\mu^3\big(\sqrt{2}\big(\big(2 + \mu^2\big)y^2 - \mu^2z^2-2\big)-4\mu z^2\big)\big]

%% file: grad_y_refined.txt
y\big[\mu^2\big(\sqrt{3}\lambda x\big(2 - 2y^2 + \mu^2z^2\big)-4\mu z^2\big(3 - 3y^2 + z^2\big)\big) - \sqrt{2}\big(2\mu^2\big( y^2-1\big)\big( 3y^2 - z^2-3\big)+\sqrt{3}\lambda\mu^3x\big( y^2-2\big)z^2\nonumber\\
&-16\big(1 - y^2 + z^2\big)^2  + \mu^4\big(y^4 + z^2\big(3 + z^2\big) - y^2\big(1 + 2z^2\big)\big)\big)\big]/\big[\mu^2\big(\sqrt{2}\big(\big(2 + \mu^2\big)y^2 - \mu^2z^2-2\big)-4\mu z^2\big)\big]